\documentclass{article}
\usepackage{spconf,amsmath,epsfig}
\usepackage{amsfonts}
\usepackage{algorithm,algorithmic}
\usepackage[font=small,skip=2pt]{caption}
\usepackage{cite}
\usepackage{graphicx,subcaption}
\usepackage{copyrightbox}
\usepackage{lipsum}

\let\OLDthebibliography\thebibliography
\renewcommand\thebibliography[1]{
	\OLDthebibliography{#1}
	\setlength{\parskip}{0pt}
	\setlength{\itemsep}{0pt plus 0.3ex}
}

\newcommand{\norm}[1]{\left\|#1\right\|}
\newcommand{\vect}[1]{\pmb{#1}}

\newcommand\numberthis{\addtocounter{equation}{1}\tag{\theequation}}

\def\bT{\ensuremath{{\bf T}}}
\def\bD{\ensuremath{{\bf D}}}
\def\bX{\ensuremath{{\bf X}}}
\def\bY{\ensuremath{{\bf Y}}}

\def\bx{\ensuremath{{\bf x}}}

\def\by{\ensuremath{{\bf y}}}
\def\bx{\ensuremath{{\bf x}}}
\def\bI{\ensuremath{{\bf I}}}

\def\be{\begin{equation}}
\def\ee{\end{equation}}
\def\ben{\begin{equation*}}
\def\een{\end{equation*}}
\def\bea{\begin{eqnarray}}
\def\eea{\end{eqnarray}}
\def\beaa{\begin{eqnarray*}}
	\def\eeaa{\end{eqnarray*}}

\title{Collaborative Sparse Priors for Infrared Image Multi-view ATR}
\name{Xuelu Li and Vishal Monga\thanks{Research supported by a grant from Lockheed Martin.}}
\address{Department of Electrical Engineering, The Pennsylvania State University}
\begin{document}
	\setlength{\belowdisplayskip}{0pt} \setlength{\belowdisplayshortskip}{0pt}
	\setlength{\abovedisplayskip}{0pt} \setlength{\abovedisplayshortskip}{0pt}
	\setlength\belowcaptionskip{-3ex}
%
\maketitle
\begin{abstract}
\vspace{0pt}
Feature extraction from infrared (IR) images remains a challenging task. Learning based methods that can work on raw imagery/patches have therefore assumed significance. We propose a novel multi-task extension of the widely used sparse-representation-classification (SRC) method in both single and multi-view set-ups. That is, the test sample could be a single IR image or images from different views. When expanded in terms of a training dictionary, the coefficient matrix in a multi-view scenario admits a sparse structure that is not easily captured by traditional sparsity-inducing measures such as the $l_0$-row pseudo norm. To that end, we employ collaborative spike and slab priors on the coefficient matrix, which can capture fairly general sparse structures. Our work involves joint parameter and sparse coefficient estimation (JPCEM) which alleviates the need to handpick prior parameters before classification. The experimental merits of JPCEM are substantiated through comparisons with other state-of-art methods on a challenging mid-wave IR image (MWIR) ATR database made available by the US Army Night Vision and Electronic Sensors Directorate.
\end{abstract}
%
%
\vspace{-8pt}
\section{Introduction}
\vspace{-5pt}
\label{sec:intro}
With developments of IR technology, ATR for IR images has attracted significant attention. Early research in ATR using IR images was focused on feature extraction from images produced through different IR sensors or the fusion of imagery \cite{GLin2015fusion,Anca2009information}. Unlike feature extraction from optical images, the choice of robust discriminative features from IR imagery remains an open problem. This is typically due to low resolution, high noise and unique physical characteristics of IR images. Nevertheless, lots of traditional algorithms for classification have been applied to IR images. Methods such as Bayesian techniques, Support Vector Machines (SVMs), Principal Component Analysis (PCA) etc. with state-of-art extracted features like Bag Of Words (BOW) and Histogram of Oriented Gradients (HOG) fall in this category\cite{Besbes2011SVM,Anca2009SVM,Mu2013BP,Khan2014HOG,Rodger2016CNN}.

\begin{figure}[t]
 	\centering
 	\includegraphics[width=7cm]{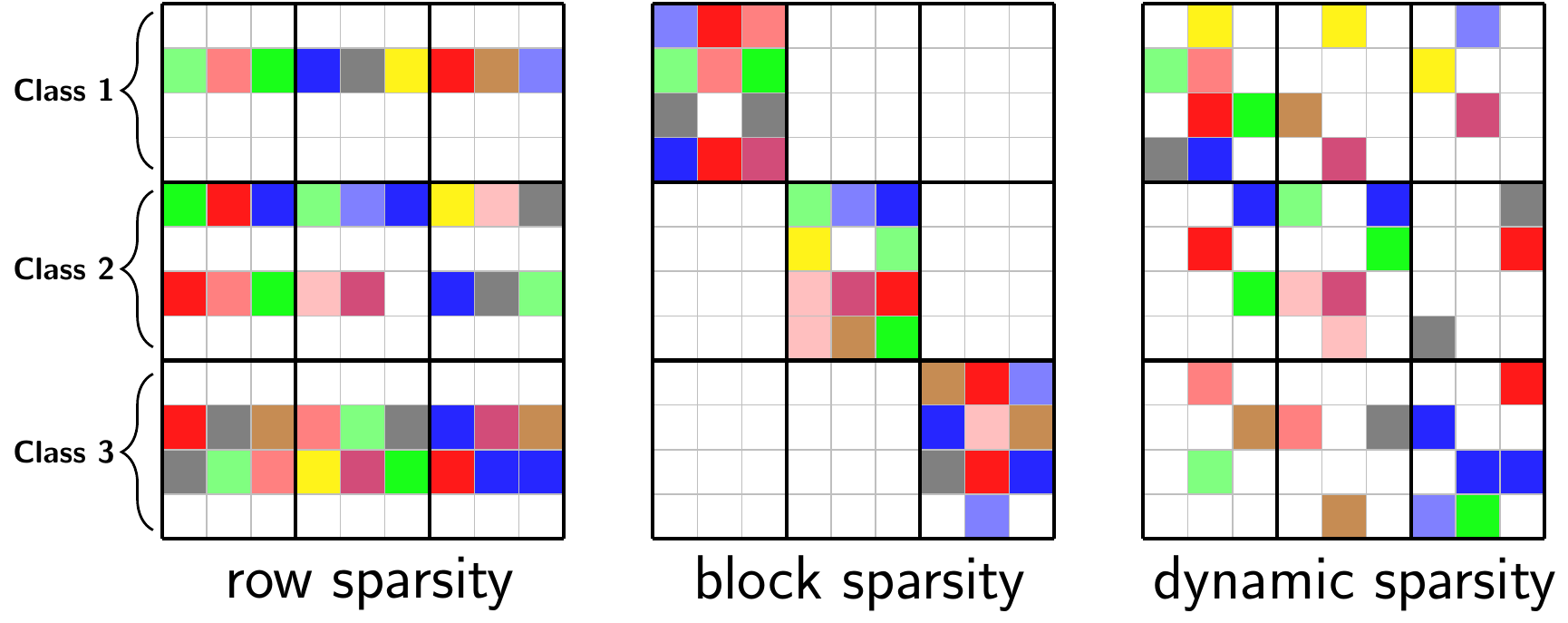}
 	\vspace{-1pt}
 	\caption{Examples structures of the sparse coefficient matrix $\mathbf X$.}
 	\label{fig:SparStruc}
 	 \vspace{-8pt}
 \end{figure}

In many cases it is possible to obtain IR images including information from different views of the same target particularly in military applications. These scenarios present the problem of multi-view ATR and recently methods have been proposed to classify such type of images effectively. Among them, the sparse representation classification (SRC) \cite{Wright2009SRC} based methods display remarkable performances by accurately recovering sparse coefficients corresponding to each class. Multi-view sparsity based classification involves a linear model where a test matrix is expanded in terms of a training dictionary of images multiplied by a coefficient matrix. The coefficient matrix is sparse but its exact structure varies by the scenario. For example, when classifying histopathological images, the row-sparsity structure is natural to maintain correspondence across different color channels\cite{Umamahesh2014SimulSpar}. However, in many other cases, sparse coefficient matrices exhibit block sparsity and dynamic sparsity \cite{Zhang2012JDSRC}, which capture more general notions of sparsity that are applicable in a wide variety of applications\cite{Hojjat2014MultiTask}. Example sparse structures of a coefficient matrix  are illustrated in Fig.\ref{fig:SparStruc}.

We address multi-view ATR via a sparsity based approach where sparse structure on the coefficient matrices is enforced by collaborative spike and slab priors. These priors have tunable parameters which can control the likelihood of each coefficient in the matrix to be active (or zero), hence allowing for more generality in the sparse structure \cite{Monga:book}. The merits of using such priors in classification have been demonstrated recently \cite{Umamahesh2015Struc}. One key challenge in practical application to classification problems is that the sparsity inducing parameter must be handpicked before the coefficient matrix is estimated. This has been done by using domain knowledge in problems such as face recognition \cite{Umamahesh2015Struc} and SONAR ATR \cite{McKay2017Sonar}. We overcome this challenge by developing a new Joint Prior and Coefficient Estimation Method (JPCEM).  Since JPCEM is also a sparsity based method, it inherits the advantages of SRC such as automatic feature discovery and resilience to noise. Experiments are carried out on a well-known database\cite{MWIRdatabase} of Mid-wave Infrared (MWIR) images collected by the US Army Night Vision and Electronic Sensors Directorate (NVESD). JPCEM is evaluated against state of the art multi-view ATR/classification methods and shown to compare favorably, particularly when training imagery is limited.
\vspace{-12pt}
\section{Sparse Representation Classification via Bayesian Framework}
\vspace{-10pt}
\subsection{Single-view Classification}
\vspace{-8pt}
\label{sec:format}
Sparse representation based classification (SRC) has become popular recently. Suppose that there are $C$ different classes, the basic relaxed form of SRC can be expressed as below:
\begin{equation} \label{eq:SRCl1}
{\bx}^\ast=\underset{{\bf x}}{\arg \min} \norm{\bx}_1 ~~s.t.~~ \norm{{\bf y-Dx}}_2<\epsilon
\end{equation}
\begin{equation} \label{eq:2}
c^\ast=\underset{c\in \{1,\dots,C\}}{\arg \min} \norm{{\by-\bD\delta_c(\bx^\ast)}}_2
\end{equation} 
where $\by\in\mathbb{R}^d$ is formed by vectorized test image, $\bD=[\bD_1,\dots,\bD_c,\dots,\bD_C]\in\mathbb{R}^{d\times K}$ is the dictionary formed by all the vectorized training images, where $\bD_c\in\mathbb{R}^{d\times{k_c}}$ is the class specific dictionary, and $\bx\in\mathbb{R}^K$ is the recovered sparse coefficient vector. $\delta_c(\bx^\ast)$ is a new vector whose only nonzero entries are the entries in $\bx^\ast$ associated with the $c^{th}$ class.
According to \cite{Derin2010Bayes}, the minimization problem in Eq.\eqref{eq:SRCl1} is equivalent to maximizing probability of observing the sparse vector $\bx$ given $\by$ assuming $\bx$ has an i.i.d Laplacian distribution in a Bayesian framework. As a result, the equation \eqref{eq:SRCl1} can also be expressed as below:
\begin{equation} \label{eq:SRCpdf}
 \bx^\ast=\underset{{\bx}}{\arg \max}f(\bx) ~~s.t.~~ \norm{{\bf y-Dx}}_2<\epsilon
\end{equation}
where $f(\bx)$ represents a sparsity inducing probabilistic prior on the coefficient vector $\bx$.

In this Bayesian set-up, we propose to work with spike and slab priors, which are known to enable fully general sparse structures \cite{Ishwaran2005Spike}: 
\begin{align*}
\by | \bD, \bx, \vect\gamma, \sigma^2 ~\sim&~ \mathcal{N}(\bD \bx, \sigma^2\bI)\numberthis\\
\bx | \vect\gamma, \lambda, \sigma^2 ~\sim &~ \prod_{i=1}^{N}\gamma_i \mathcal{N}(0,\sigma^2\lambda^{-1}) + (1-\gamma_i)\mathbb{I}(x_i=0)\\
\vect\gamma | \vect\kappa~\sim &~ \prod_{i=1}^{N}\mbox{Bernoulli}(\kappa_i)
\end{align*}
where $\mathcal{N}(\cdot)$ represents the normal distribution, $\vect\gamma$ is the indicator variable for vector $\bx$, i.e., $\gamma_i = 0$ if $x_i = 0$, otherwise $\gamma_i=1$, and $\kappa_i$ is the probability of $\gamma_i$ being $1$. Maximum-a-posteriori (MAP) estimation using such priors are known \cite{Monga:book} to lead to the following optimization problem:
 \begin{equation}\label{eq:Costfun}
 (\bx^\ast, \vect\gamma^\ast)=\underset{{\bx,\vect\gamma}}{\arg \min}(\norm{\by-\bD\bx}_2^2+\lambda\norm{\bx}^2_2+\sum_{i=1}^N\gamma_i\rho_i
 \end{equation}
 where $\rho_i=\sigma^2\log(\frac{2\pi\sigma^2(1-\kappa_i)^2}{\lambda\kappa^2_i})$. The parameters $\kappa_i$ in Eq.\eqref{eq:Costfun} have different values for each sparse coefficient, which is essential to classification under such a Bayesian framework. It
 is noted that this is a more general formulation than
 the framework in \cite{Yen2011MMA} where authors simplified
 the optimization problem by assuming the same $\kappa_i$ for each
 coefficient and the last term in Eq.(\ref{eq:Costfun}) reduces to $\rho \Vert \mathbf x \Vert_{0}$. The optimization problem in Eq.(\ref{eq:Costfun}) (for fixed and known $\rho_i$) is challenging with no-known global minima. Nevertheless, sub-optimal algorithms have been developed recently etc \cite{Hojjat2014MultiTask,Hojjat2015ICR,Tiep2017AMP}. 
 \vspace{-5pt} 
\subsection{Multi-view classification}
\vspace{-3.5pt}
For the situations where multiple measurements from different views of the same target are obtained, the model described above can also be well applied. In such situations, collaborative spike and slab priors can be written as:
\begin{align*}
\bY | \bD, \bX,  \vect\Gamma, \sigma^2 ~\sim \prod_{m=1}^{M}\mathcal{N}&(\bD \bx_m, \sigma^2\bI) \numberthis \label{eq: 10}\\
\bX | \vect\Gamma, \lambda, \sigma^2 ~\sim \prod_{m=1}^{M}\prod_{i=1}^{N} ~\gamma_{m_i}\mathcal{N}&(0,\sigma^2\lambda^{-1}) + (1-\gamma_{m_i})\mathbb{I}(x_i=0)\\
\vect\Gamma | {\bf K}~\sim\prod_{m=1}^{M}\prod_{i=1}^{N}& ~\mbox{Bernoulli}(\kappa_{m_i})
\end{align*}
and the corresponding optimization problem will become:
\begin{align*}\label{eq: 11}
&(\bX^\ast, \vect \Gamma^\ast)=\numberthis
\\&\underset{{\bX,\vect\Gamma}}{\arg \min}~{\sum_{m=1}^M(\norm{\by_m-\bD\bx_m}_2^2+\lambda\norm{\bx_m}^2_2+\sum_{i=1}^N\gamma_{m_i}\rho_{m_i})}
\end{align*}
where $\rho_i=\sigma^2\log(\frac{2\pi\sigma^2(1-\kappa_{m_i})^2}{\lambda\kappa^2_{m_i}})$, $\bX=[\bx_1,...,\bx_m,...,\bx_M]$ $\in\mathbb{R}^{K\times M}$, $\vect\Gamma=[\vect\gamma_1,\dots,\vect\gamma_m,\dots,\vect\gamma_M]\in\mathbb{R}^{K\times M}$, $\bD=[\bD_1,...,\bD_c,$ $...,\bD_C]\in\mathbb{R}^{d\times K}$,  $\bD_c=[\bD^1_c,...,\bD^m_c,...,\bD^M_c]\in\mathbb{R}^{d\times k_c}$, $\bD^m_c\in\mathbb{R}^{d\times t_m}$,  and $\sum_{m=1}^{M}t_m=k_c$ ($t_m$ is the number of training samples from the $m^{th}$ view of a target). 
  Since $\kappa_{m_i}$ is assumed to have different values for each coefficient corresponding to each view, the framework is capable of representing general sparse structures of the matrix $\bX$. Similar to the single-view problem, after solving the optimization problem corresponding to each class, the class label can be decided as below:
  \begin{equation} \label{eq:Classification}
  c^\ast=\underset{c\in \{1,\dots,C\}}{\arg \min} \sum_{m=1}^{M}\norm{{\by_m-\bD\delta_c(\bx^\ast_m)}}_2
  \end{equation}
  \vspace{-8pt}
\section{Joint Prior and Coefficient Estimation}
\vspace{-8pt}
\subsection{The optimization problem}
\vspace{-4pt}
\label{sec:pagestyle}
The collaborative Bayesian framework based classification described above shows us that estimating values of the parameter $\kappa_{m_i}$ accurately is key to obtaining high classification accuracy. Unfortunately, as the parameter that induces the sparsity, the values of $\kappa_{m_i}$ are different for each coefficient each view.  Generally, we can conduct cross-validation or use the traditional method MCMC mentioned in \cite{Umamahesh2015Struc} to estimate the values. However, both the methods require daunting calculation load, which becomes especially serious when the number of classes and views become quite large. Therefore, in this paper, we propose a new Joint Prior and Coefficient Estimation Method (JPCEM) to estimate the parameter $\kappa_{m_i}$ more effectively. In particular, we propose a novel solution for an extended version of the problem in Eq.\eqref{eq:Costfun}, where the $\kappa_i$'s and hence $\rho_i$'s are determined automatically in an iterative procedure as opposed to be fixed and known.
  \begin{figure}[t]
  	\centering
  	\includegraphics[width=6.5cm]{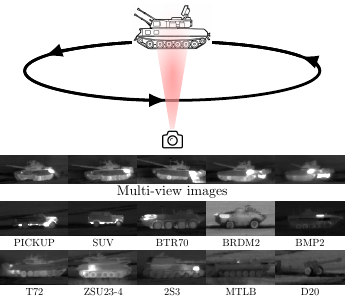}
  	\caption{database capturing and data samples}
  	\label{fig: sample}
  	\vspace{-4pt}
  \end{figure}

We begin by observing that $\rho_{m_i}$ is monotonically decreasing with respect to $\kappa_{m_i}$, and most values of $\rho_{m_i}$ will be higher than $0$ if the value of $\frac{\sigma^2}{\lambda}$ is chosen suitable (an example is shown in the experiments). That is, the higher the probability that a sparse coefficient is not zero, the lower $\rho$ will be. As a result, we can propose a reasonable assumption that the larger the absolute value of the sparse coefficient, the higher the probability that the coefficient will not be zero. On the basis of this assumption,  $\rho_{m_i}$ can be expressed as
\begin{equation}
\rho_{m_i}=\sigma^2\log(\frac{2\pi\sigma^2(1-\left|\frac{x_{m_i}}{(1+\alpha)x_{M}+\epsilon}\right|)^2}{\lambda(\left|\frac{x_{m_i}}{(1+\alpha)x_{M}+\epsilon}\right|)^2})
\end{equation}
where $x_M$ is the maximum value in the vector $\bx_m$, $\epsilon$ is a small value bigger than 0, and $\alpha$ is chosen once the value of $\frac{\sigma^2}{\lambda}$ is set and fixed to promise the condition that $\rho_{m_i}>0$, we focus on solving the following problem:
 \begin{equation}\label{eq:Reweight}
 \bX^\ast=\underset{{\bX}}{\operatorname{argmin}}\sum^M_{m=1}({\norm{{\bf \hat{y}}_m-\bf{\hat{D}}{\bf x}_m}_2^2+\norm{{\bf w}_m \circ |\bx_m|}_1})
 \end{equation}
 where $\circ$ represents the Hadamard product,  ${\bf \hat{y}_m}=[\by_m^\bT~~\bf{0}]^T$, $\bf{\hat{D}}=[\bD^\bT ~\bf{\sqrt{\lambda}\vect I}]^\bT$, and $w_{m_i}=\frac{\rho_{m_i}}{\left|x_{m_i}\right|+\epsilon}$, $\epsilon>0$. 
 To tie the problem in Eq.\eqref{eq:Reweight} to the one in Eq.\eqref{eq:Costfun} note that the term $\norm{{\bf w}_m \circ |\bx_m|}_1$ approximates or mimics $\sum_i \gamma_i \rho_i$. In particular, when $\gamma_{m_i} = 1$ and hence $|x_{m_i}|\ne0$, $\omega_{m_i}|x_{m_i}|=\frac{\rho_{m_i}}{\left|x_{m_i}\right|+\epsilon} |x_{m_i}|\rightarrow\rho_{m_i}$, i.e.\ $\rho_i$ is updated. Similarly, when $\gamma_i = 0$ and hence $|x_{m_i}|=0$, the update generates a $0$. 

If ${w_m}_i$ is inversely related to the magnitude of $x_{m_i}$, it can be regarded as a standard re-weighted-$l_1$ form, which is true  since except for $\frac{1}{|x_{m_i}|}$, $\rho_{m_i}$ is also monotonically decreasing with respect to $x_{m_i}$. The entire sparse matrix $\mathbf X$ is obtained column-wise by solving $M$ problems (one for each view) of the form in Eq.\eqref{eq:Reweight}. 
  \vspace{-15pt}
 \subsection{The iterative algorithmic procedure}
  \vspace{-7pt}
 Our overall solution is described in Algorithm \ref{algorithm1}. The essence of our method is to start with initial values of parameters and solve a sequence of convex problems, such that each problem in the sequence is a re-weighted $l_1$ problem. Solutions of standard re-weighted $l_1$ problems can be found in \cite{Candes2008RWl1}. The updates to  $\kappa_{m_i}, \rho_{m_i}, w_{m_i}$ are done iteratively and the updated parameters are used to form the next problem in the sequence.
 
 Finally, the class label can be decided according to the sum of residues from different views as in Eq.\eqref{eq:Classification}.
 \begin{algorithm}[t]
 	\renewcommand{\algorithmicrequire}{\textbf{Input:}}
 	\renewcommand{\algorithmicensure}{\textbf{Output:}}
 	\caption{JPCEM}
 	\label{alg:MC}
 	\begin{algorithmic}
 		\label{algorithm1}
 		\REQUIRE $\bD$, $\bY=[\by_1,...,\by_M]$
 		\STATE Initialize: $\epsilon$, $\alpha$, $\sigma$, $\lambda$, $w_{m_i}^{(0)}=1$, $\kappa_{m_i}^{(0)}=0.5$, $i = 1,...,N$, $\bX^{(0)}=[\bx_1^{(0)},...,\bx_M^{(0)}]={\bf 0}$, $\bX^{(1)}=[\bx_1^{(1)},...,\bx_M^{(1)}]={\bf 1}$, $k=1$.
 		\WHILE{$\norm{\bx^{(k)}_m-\bx^{(k-1)}_m}^2>\epsilon$ for $m\in[1,..,M]$}
 		\STATE 
 		\vspace{-5pt}
 		\begin{equation*}
 		\bx^{(k)}_m=\underset{{\bX_m}}{\operatorname{argmin}}~{\norm{{\bf\hat{y}}_m-{\bf\hat{D}}{\bf x}_m}_2^2+\norm{{\bf w}_m \circ \left|{\bf x}_m\right|}_1}
 		\end{equation*}
 		\vspace{-5pt}
 		\STATE Update\\
 		$\kappa_{m_i}^{(k)}=\left|\frac{x_{m_i}^{(k)}}{(1+\alpha)x_M^{(k)}+\epsilon}\right|$, $\rho_{m_i}^{(k)}=\sigma^2\log(\frac{2\pi\sigma^2(1-\kappa_{m_i}^{(k)})^2}{\lambda(\kappa_{m_i}^{(k)})^2+\epsilon})$, $w_{m_i}^{(k)}=\frac{\rho_{m_i}^{(k)}}{\left|x_{m_i}^{(k)}\right|+\epsilon}$, $k=k+1$
 		\ENDWHILE
 		\ENSURE $ \bX^\ast=[\bx_1^{(k)},...,\bx_M^{(k)}]$, $\gamma_{m_i}^\ast=\frac{x_{m_i}^{(k)}}{\left|x_{m_i}^{(k)}\right|+\epsilon}$
 	\end{algorithmic}
 \end{algorithm}  
 \vspace{-10pt}
\section{Experimental Results}
\vspace{-10pt}
\label{sec:experiments}

The proposed JPCEM method is evaluated on the well-known MWIR database collected by NVESD, and the following state-of-art methods are selected specifically to be compared against: Graph-based multi-view classification method (GMCM)\cite{Kokiopoulou2010Graph}, which can be considered among the most effective non-SRC-based multi-task classification method; Joint SRC (JSRC) method\cite{Zhang2012JSRC}, which is suitable for the situations where row sparsity property can hold; and Joint Dynamic SRC (JDSRC) method\cite{Zhang2012JDSRC}, which is suitable for more general dynamic sparse structures.

\begin{figure}[t]
		\centering
		\includegraphics[width=8.7cm]{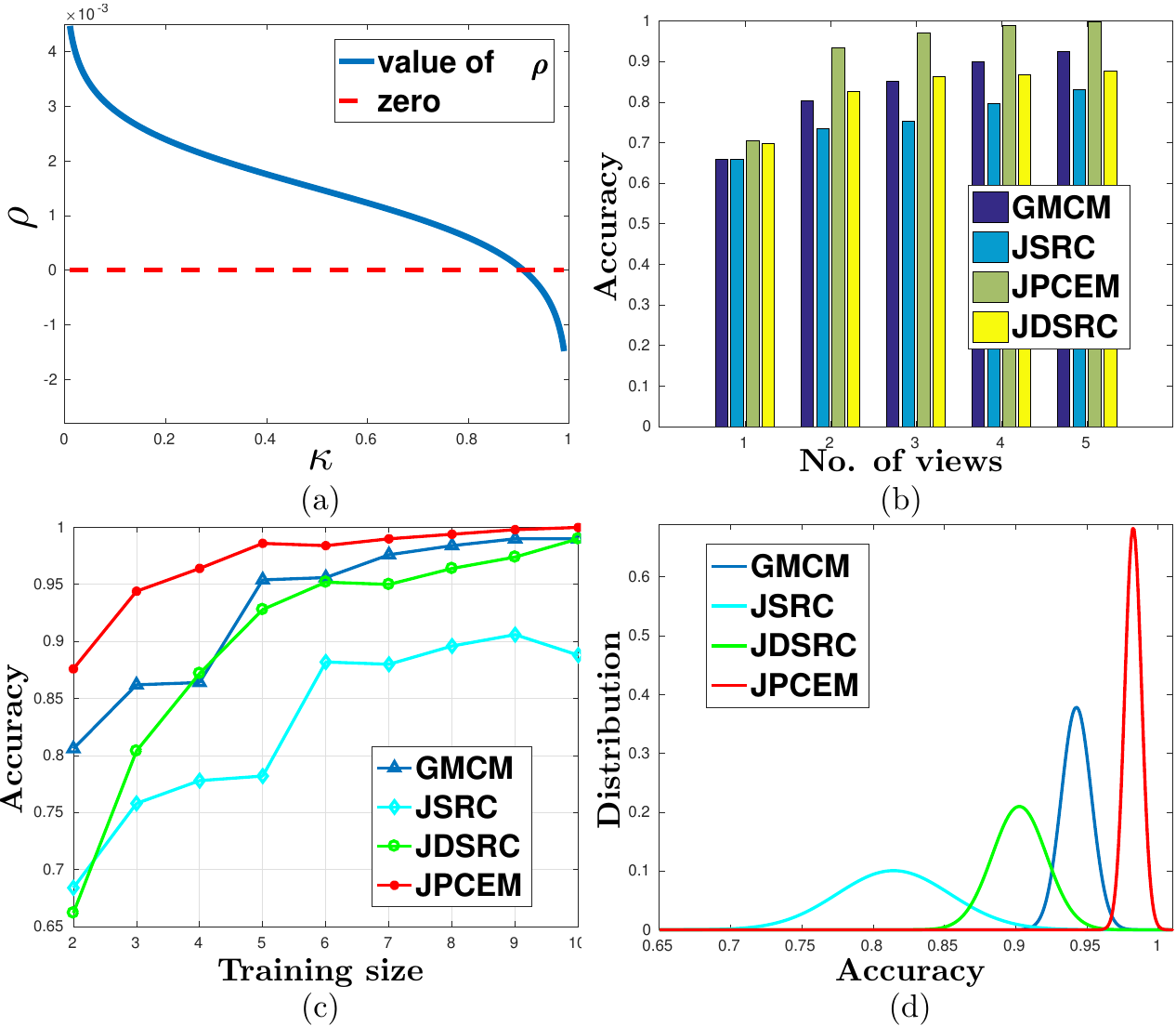}
	
	\caption{Results: (a) $\rho$ vs. $\kappa$, (b) Classification accuracy vs. no. of views, (c) Classification accuracy vs. no. of training samples, (d) Exploring Selection Bias: Distribution of classification rates over 20 different random selection of the {\em same number} of training samples.}
	\label{fig:experiment}
	\vspace{-5pt}
\end{figure}

{\bf MWIR data base}: The database is collected by NVESD to support the ATR algorithm development community\cite{MWIRdatabase}. It contains MWIR images of 10 different non-human targets including civilian vehicles, military vehicles, carriers and weapons (carried on the vehicles or carriers). The images are captured by letting the target run on circles with different ranges and distances from a camera at the same location. As a result, the images from different views of the targets are available immediately. The process and example images are illustrated in Fig.\ \ref{fig: sample}. In our experiment, we extract \footnote{See \cite{PreProcessing} for the extraction process.} the part including the target from the available images and rescale the cropped image into the size of $40\times20$ pixels. 127 images per view of each target class are used as training images, and 50 images per view per class are used as test images. 
The final classification results from the experiments under different conditions can be seen in Fig.\ref{fig:experiment} (set $\alpha=\frac{1}{9}$). Fig.\ref{fig:experiment}(a) shows the relation between $\kappa_{m_i}$ and $\rho_{m_i}$ by setting $\sigma=0.018$ and $\lambda=0.00002$. As expected, $\rho_{m_i}$ and $\kappa_{m_i}$ are inversely related with a monotonic decrease of one w.r.t the other. Fig.\ref{fig:experiment}(b) shows the classification accuracy by changing the number of views from 1 to 5 (training size: 5 images per view per class). Fig.\ref{fig:experiment}(c) shows the classification accuracy by changing the training size from 2 to 10 images per view per class (number of views fixed at 5). Fig.\ref{fig:experiment}(d) shows a Gaussian fit to the histogram of classification accuracy values across 20 different ways of selecting training samples (5 views, training size: 5 images per view per class). Remarkably in Fig.\ \ref{fig:experiment}(d), JPCEM achieves the highest mean value while exhibiting the smallest variance indicating not only high classification accuracy but robustness to the exact choice of training samples.
\vspace{-14pt}
\section{Conclusion}
\vspace{-11pt}
We propose a sparsity constrained framework for automatic classification and recognition of IR images. Our method crucially alleviates the need to know sparsity inducing model parameters in advance, which are in turn jointly estimated with sparse coefficients. This adaptation to the underlying data set results in vastly improved classification results.

\vspace{-7pt}
\small
\bibliographystyle{IEEEbib}
\bibliography{strings,refs}
\end{document}